\documentclass[a4paper]{article}
\usepackage{amsmath,amssymb,amsfonts,bbm}
\usepackage{hyperref}
\usepackage{graphicx}
\usepackage[margin=3cm]{geometry}

\usepackage{caption}
\numberwithin{equation}{section}

\newcommand{\id}{{\boldsymbol{\mathbbm{1}}}}
\DeclareMathOperator{\tr}{tr}
\DeclareMathOperator{\sym}{sym}
\DeclareMathOperator{\dev}{dev}
\DeclareMathOperator{\skewp}{skew}

\DeclareMathOperator{\Curl}{Curl}
\DeclareMathOperator{\grad}{grad}

\DeclareMathOperator{\polar}{polar}
\DeclareMathOperator{\sech}{sech}

\newcommand{\B}{\mathbf}

\begin{document}
\title{Chirality in the plane}
\author{\renewcommand{\thefootnote}{\arabic{footnote}}
  Christian G. B\"ohmer\footnotemark[1] \ and
  Yongjo Lee\footnotemark[2] \ and
  Patrizio Neff\footnotemark[3]
}

\date{\today}

\footnotetext[1]{Christian G. B\"ohmer, Department of Mathematics, University College London, Gower Street, London, WC1E 6BT, UK, email: c.boehmer@ucl.ac.uk}

\footnotetext[2]{Yongjo Lee, Department of Mathematics, University College London, Gower Street, London, WC1E 6BT, UK, email: yongjo.lee.16@ucl.ac.uk}

\footnotetext[3]{Patrizio Neff, Fakult\"at f\"ur Mathematik, Universit\"at Duisburg-Essen, Thea-Leymann-Stra\ss e 9, 45127 Essen, Germany, email: patrizio.neff@uni-due.de} 

\date{\today}
\maketitle

\begin{abstract}
It is well-known that many three-dimensional chiral material models become non-chiral when reduced to two dimensions. Chiral properties of the two-dimensional model can then be restored by adding appropriate two-dimensional chiral terms. In this paper we show how to construct a three-dimensional chiral energy function which can achieve two-dimensional chirality induced already by a chiral three-dimensional model. The key ingredient to this approach is the consideration of a nonlinear chiral energy containing only rotational parts. After formulating an appropriate energy functional, we study the equations of motion and find explicit soliton solutions displaying two-dimensional chiral properties.
\end{abstract}

\mbox{}

\textbf{Keywords:} chiral materials, planar models, Cosserat continuum, isotropy, hemitropy, centro-symmetry

\mbox{}

\textbf{AMS 2010 subject classification:} 74J35, 74A35, 74J30, 74A30

\mbox{}

\section{Introduction}

\subsection{Background}

A group of geometric symmetries which keeps at least one point fixed is called a point group. A point group in $d$-dimensional Euclidean space is a subgroup of the orthogonal group $\mathrm{O}(d)$. Naturally this leads to the distinction of rotations and improper rotations. Centrosymmetry corresponds to a point group which contains an inversion centre as one of its symmetry elements. Chiral symmetry is one example of non-centrosymmetry which is characterised by the fact that a geometric figure cannot be mapped into its mirror image by an element of the Euclidean group, proper rotations $\text{SO}(d)$ and translations.

This non-superimposability (or chirality) to its mirror image is best illustrated by the left and right hands: there is no way to map the left hand onto the right by simply rotating the left hand in the plane. This  geometrical feature of chirality can be found in many molecules which can have distinct chemical properties. A fairly stable or harmless substance can have an unstable or noxious substance as its chiral counterpart~\cite{SS2009}.

If one applies a Lorentz boost to a particle with spin in its momentum direction in one frame of reference while retaining its spin direction, this will cause the opposite direction of momentum to another frame of reference. This discrete symmetry is known as parity and leads to the notion of left or right-handedness in particle physics, similar to chirality.

Elasticity theories with microstructure contain nine additional degrees of freedom which consist of three micro-rotations, one micro-volume expansion and five micro-shear deformations. If we restrict these microdeformations to be rigid, one deals with 3 additional degrees of freedom, the microrotations. The resulting model is often referred to as Cosserat elasticity and was pioneered by the Cosserat brothers~\cite{EC} as early as 1909 fully in its geometrically nonlinear setting. The more general micromorphic model was developed by Eringen in~\cite{AE1964-1, AE1964-2, AE1}, for more recent developments the reader is referred to~\cite{DL2001, PN2004, PN2006, PN2007, SF2009, PN2013, PN2014, PN2016, PN2017,NFB:2019}.

In continuum mechanics, it is often observed that chiral materials in three-dimensional space when projected into the two-dimensional plane, lose their chirality~\cite{RL1982,AS2011,XL2012-1}. A linear energy function (quadratic energy in small strains) for an isotropic material in the centro-symmetric case was studied in~\cite{RM1964} and the absence of odd-rank isotropic tensors implicitly implied the lack of material parameters for the non-centrosymmetric case. Similar works~\cite{RL1982, RL2001-1} considered an energy function which contained a fourth order isotropic tensor with chiral coupling terms by identifying axial tensors as being asymmetric under the inversion. When one attempts to apply these ideas to the planar case, the chiral coupling term turns out to vanish~\cite{AS2011, XL2012-1} and one arrives at an isotropic (and centrosymmetric) model without chirality. A rank-five isotropic tensor was introduced in~\cite{SP2011} to impose chirality on the energy containing a single chiral material parameter. This type of chirality was related to the gradient of rotation, which led to the existence of torsion. Based on the assertion that hyperelastic Cosserat materials are hemitropic ($\mathrm{SO}(3)$-right-invariant) if and only if the strain energy is hemitropic, a set of hemitropic strain invariants was given in~\cite{KC1981}. Many attempts were made to understand the mechanism behind the loss of chirality and in constructing a generic two-dimensional chiral configuration without referring to higher dimensions.

A chiral rank-four isotropic tensor was used in~\cite{XL2012-1} to derive a chiral material constant in the equations of motion, and in subsequent works~\cite{XL2012-2, XL2016} the two-dimensional chirality problem is further considered. A planar micropolar model is proposed in~\cite{YC2014-1} with the help of the irreducible decomposition of group representation. Recent growing interests of planar chirality~\cite{VF2006, AP2003, WZ2006, MX2012, ZL2013} concern the polarised propagation of electromagnetic waves. The optical behaviour indicates that planar chirality behaves differently from its three-dimensional counterpart. In~\cite{OA2016} a two-dimensional chiral optical effect in nanostructure is studied in comparison with three-dimensional chirality. The two-dimensional micropolar continuum model of a chiral auxetic lattice structure in connection with negative Poisson's ratio is discussed in~\cite{AS2011, DP1997, YC2014-1, YC2014-2}. The theoretical analysis of planar chiral lattices is compared with experimental results in~\cite{WZ2018}. A schematic description of chiral transformations and the changes of the number of symmetry groups from higher dimensions (macroscale chiral layers) to lower dimensions (molecules of chiral line structure) is outlined in~\cite{RR2009}, backed by various experimental results. Further developments in three-dimensional chiral structures can be found in~\cite{FC2016, DI2017, WW2018}.

\subsection{Principle aim}

Let us begin with an immediate observation regarding the rotational field. In three-dimensional space it is in general non-Abelian while it becomes Abelian in two dimensions. So, a certain loss of information is expected when projecting to a lower dimensional space. In this paper we will construct a new geometrically nonlinear energy term which is explicitly chiral in three dimensions and which does not loose this property when applied to the planar problem.

Recall the first planar Cosserat problem. The displacement vector is given by $\B{u}=(u_1,u_2,0)$ while planar rotations are described by a rotation axis $\B{a}=(0,0,a_3)$. Then the dislocation density tensor $\overline{K}=\overline{R}^T\Curl\overline{R}$ (the overline indicates quantities with microstructure) is non-zero. The microrotations are described by the orthogonal matrix $\overline{R}$. It does contain a $2\times 2$ zero block matrix in the planar indices. This induces several orthogonal relations, for example, with the first Cosserat deformation tensor $\overline{U}=\overline{R}^TF$, here $F$ is the deformation gradient, which make it impossible to construct coupling terms in the energy which do not vanish identically in the plane, see the detailed discussions in~\cite{HJ2011, CB2017-1}. In our previous paper~\cite{CB2017-1} we constructed a generic two-dimensional theory with chirality without reference to a higher-dimensional model. We also speculated that `it might be possible to construct chiral terms using non-linear functionals beyond the usual quadratic terms which yield a non-trivial planar theory'. We are now able to answer this question affirmatively by constructing a rather simple energy term which displays chirality in three \emph{and} two dimensions. 

\begin{figure}[!htb]
  \center{\includegraphics[scale=0.3]{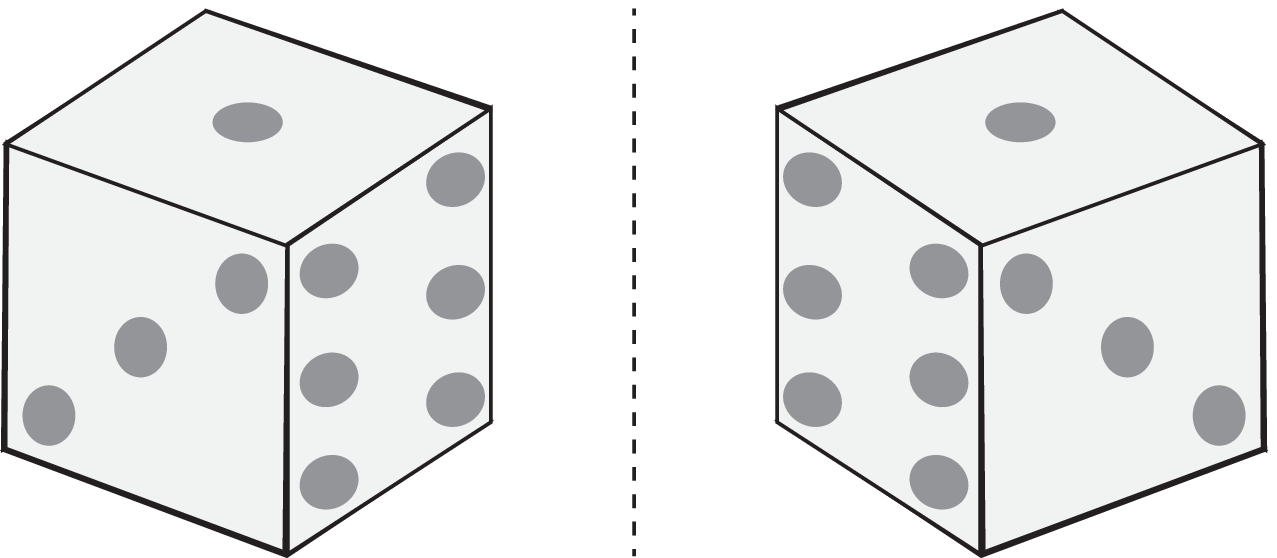}}
  \center{\includegraphics[scale=0.3]{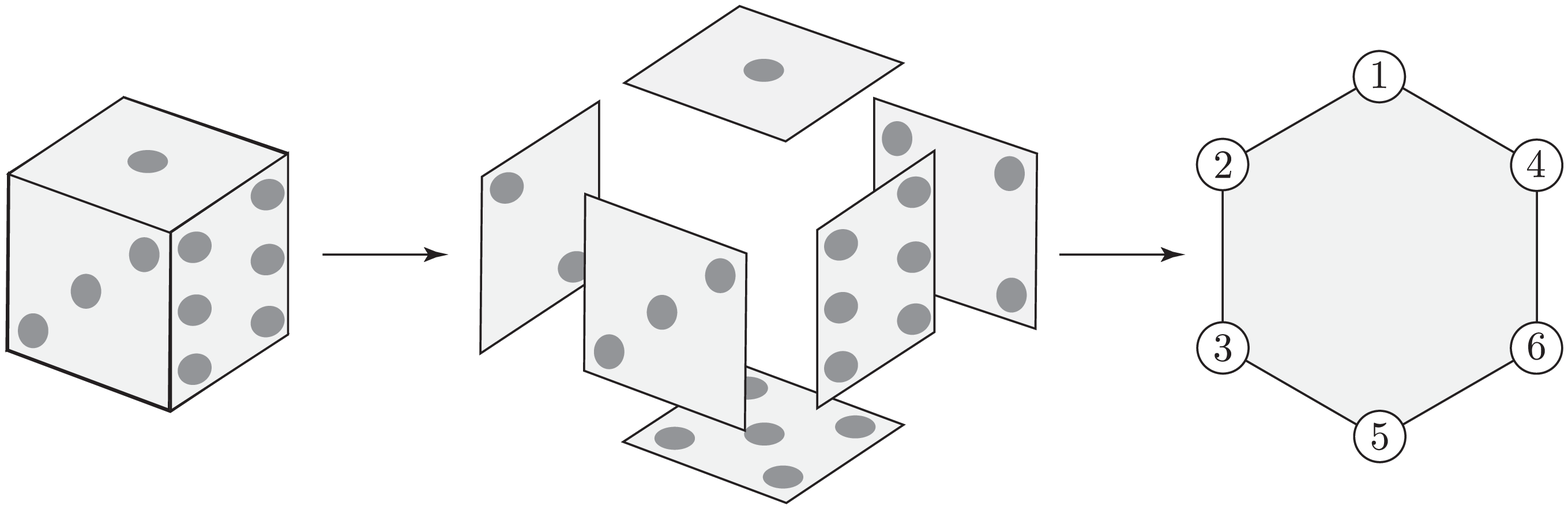}}
  \center{\includegraphics[scale=0.3]{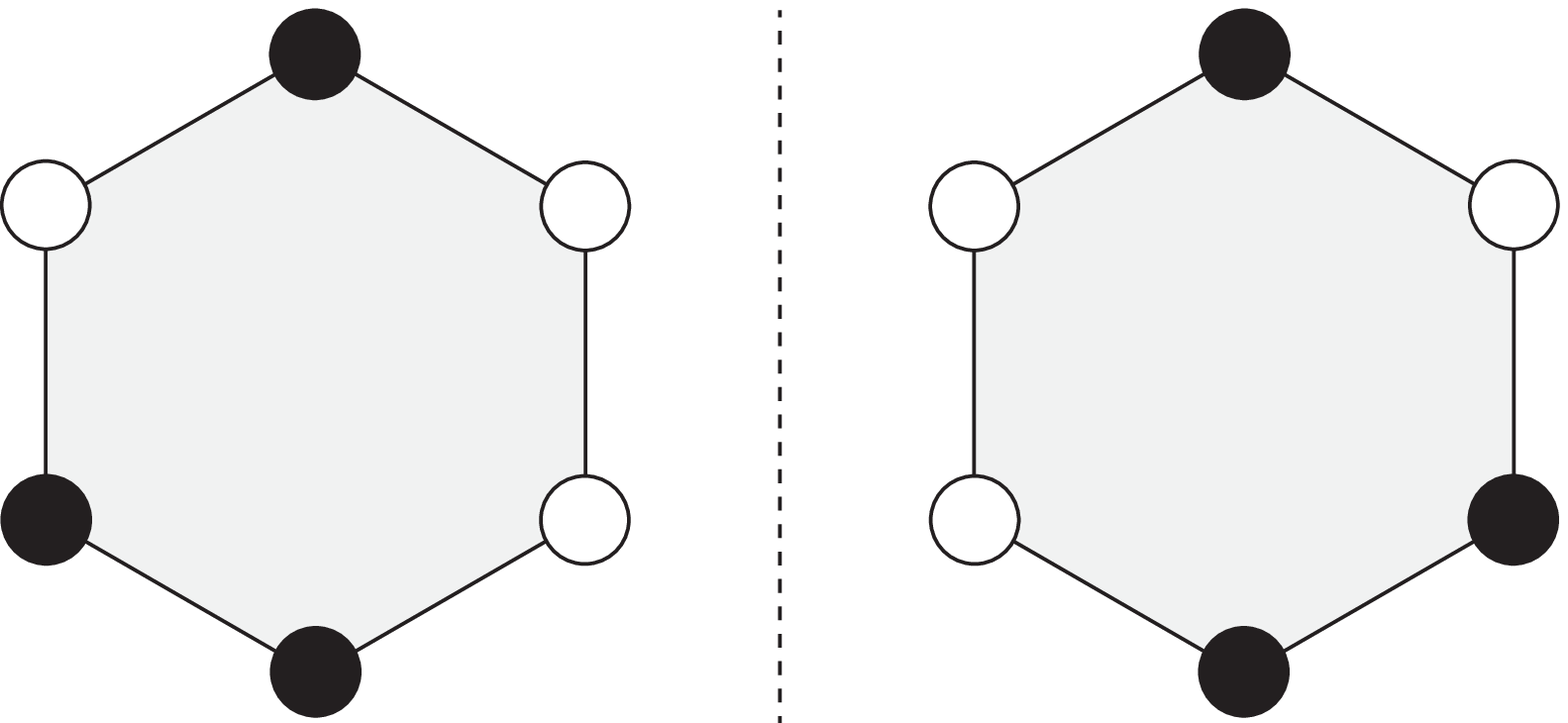}}
  \caption{We project the image of a chiral die into a two-dimensional plane along with a particular point of view to obtain a regular hexagon with numbered balls on its vertices where the numbers are exactly in the order as printed on the given die. Further, we map even numbers into white balls and odd numbers into black balls. We get a chiral hexagonal two-dimensional structure.}
\label{f001}
\end{figure}

Let us illustrate a simple example regarding chirality in three dimensions which can be translated into two dimensions, see Fig.~\ref{f001}. 
This example indicates, at least at an intuitive level, that one can construct a three-dimensional chiral structure and a projection such that the resulting two-dimensional space inherits the chirality in a certain way. However we must note that if we had a different way of labelling dots on the chiral die we would not find a chiral hexagon on the plane. In other words, the chirality preservation through the projection into the lower-dimensional space depends critically on its construction.

The remainder of this paper is organised as follows: Section~\ref{sec:chiraldef} explains our definition of chirality and its application to the deformation tensor and the dislocation density tensor. Section~\ref{sec:chiralterm} shows the construction of one possible chiral term which can form the basis for constructing a multitude of other chiral terms. In Section~\ref{sec:eom} we explicitly state the dynamic equations of motion in the plane and show that chirality does not vanish. We then solve those equations and explain their chiral structure. A discussion of results ends the paper with Section~\ref{sec:dis}.

\section{Chiral energy terms}

\subsection{Defining chirality}
\label{sec:chiraldef}

We define the coordinate-inversion operator $\#$ so that it acts on a function $\varphi=\varphi(x,y,z)$
\begin{align}
  \label{2.1}
  \varphi^\#(x,y,z):=\varphi(-x,-y,-z)\;.
\end{align}
This means we evaluate the function $\varphi$ at the inverted coordinates $(-x,-y,-z)$. On the deformation gradient $F=F(x,y,z)=\nabla\varphi(x,y,z) = \id + \nabla \mathbf{u}(x,y,z)$, this operator acts as
\begin{align}
  \label{2.2}
  F^\#=(\nabla\varphi)^\#=-\nabla\varphi(-x,-y,-z)=-F(-x,-y,-z)\;.
\end{align}
i.e.~under the operator $\#$, $F$ picks up the negative sign at the inverted coordinates. The stretch part of $F$, defined by $U=\sqrt{F^TF}$, is invariant under $\#$ since
\begin{align}
  \label{2.3}
  U^{\#}=\sqrt{(F^TF)^{\#}}=\sqrt{((F^T)^\#\;F^\#)}=\sqrt{F^TF}=U\;.
\end{align}
We note that the operation $\#$ when acting on a product of matrices does not reverse the order of the multiplication.

On the other hand, to see the action of the operator $\#$ on the polar part of $F$ in the polar decomposition \cite{PN2014-1} $F=R\,U$, $\polar(F)=R$ we consider the action on the polar decomposition of $F$,
\begin{align}
  \label{2.4}
  F^{\#}=(R\,U)^{\#}=R^{\#}\,U^{\#}=R^{\#}\,U=-F(-x,-y,-z)\;.
\end{align}
This implies that the orthogonal matrix $R=\polar(F)$ transforms under $\#$ as
\begin{align}
  \label{2.5}
  R^\#(x,y,z)=-R(-x,-y,-z)
\end{align}
in complete analogy to the transformation of $F$. Consequently,
\begin{align}
  \label{2.6}
  (R^\#)^T=-(R^T)(-x,-y,-z)=(R^T)^\#\;,
\end{align}
which means that $F$ (or $F^T$) and its polar part $R$ (or $R^T$) transform in the same way under $\#$. The transformation properties of a general rotational matrix under inversion are discussed in greater detail in Appendix~\ref{app:rot}. Now, consider the row-wise curl of $R^\#$, see~\cite{PN2008}, which reads
\begin{align}\label{2.7}
  \Curl (R^\#)=
  \begin{pmatrix}
    \partial_yR^\#_{xz}-\partial_zR^\#_{xy}&\partial_zR^\#_{xx}-\partial_xR^\#_{xz}&\partial_xR^\#_{xy}-\partial_yR^\#_{xx}\\
    \partial_yR^\#_{yz}-\partial_zR^\#_{yy}&\partial_zR^\#_{yx}-\partial_xR^\#_{yz}&\partial_xR^\#_{yy}-\partial_yR^\#_{yx}\\
    \partial_yR^\#_{zz}-\partial_zR^\#_{zy}&\partial_zR^\#_{zx}-\partial_xR^\#_{zz}&\partial_xR^\#_{zy}-\partial_yR^\#_{zx}
  \end{pmatrix}\;.
\end{align}
We observe that each partial derivative gives additional minus signs to the matrix elements of $R^\#$ using the chain rule, for instance
\begin{align}\label{2.8}
  \partial_y(R^\#)=-\partial_y R(-x,-y,-z)=\partial_{-y}R(-x,-y,-z)=(\partial_yR)(-x,-y,-z)
\end{align}
so that the partial derivative of $R$ with respect to $y$ is evaluated in the usual Cartesian coordinates but the quantity $\partial_yR$ is viewed in the inverted coordinate system $(-x,-y,-z)$. Then 
\begin{align}
  \label{2.9}
  \Curl (R^\#)(x,y,z)=(\Curl R)(-x,-y,-z)\;.
\end{align}
Hence, we see that the operator $\Curl$ negates the minus sign obtained when $\#$ is acted on $R$ but we can easily construct a curvature measure which transforms like (\ref{2.4}), such as
\begin{align}
  \label{2.10}
  (R^T\Curl R)^\# = (R^\#)^T\Curl R^\# =-(R^T\Curl R)(-x,-y,-z) \;.
\end{align}
It is no coincidence that the dislocation density tensor is exactly of this form. Clearly, we will arrive at the same conclusions for $\overline{R}=\polar(\overline{F})$ and some arbitrary $\overline{F} \in \mathrm{GL}(3)$
\begin{align}
  \overline{F}^\#(x,y,z)&=-\overline{F}(-x,-y,-z)\;,\\
  \overline{R}^\#(x,y,z)&=-\overline{R}(-x,-y,-z)\;,\\
  \overline{K}^\#=(\overline{R}^T\Curl\overline{R})^\#(x,y,z)&=-(\overline{R}^T\Curl\overline{R})(-x,-y,-z)=-\overline{K}\;.
  \label{2.11}
\end{align}

In this way, we would like to investigate some frequently appearing matrix quantities in the Cosserat elasticity energy functional, accompanied by conventional matrix operations such as transpose, trace or Frobenius scalar product. We will see whether they are chiral or not. This will lead to a simple combination of products which yields a chiral energy functional.

\subsection{Objective and chiral energy functional}
\label{sec:chiralterm}

In searching for a generic objective~\cite{IM2018} and chiral term for a chiral energy functional, we begin by recalling objectivity. We call an energy functional objective if it is invariant under global (macroscopic) left-rotations $Q$ with $\det Q=+1$, here $Q \in \mathrm{SO}(3)$ is a constant orthogonal matrix. Under the action of $Q$ we have for $F$ and $\overline{R}$ respectively
\begin{equation}
  \label{2.14}
  F = R\,U \rightarrow QR\,U = QF\;,\qquad  \overline{R} \rightarrow Q \,\overline{R} \;.
\end{equation}
One verifies that the dislocation density tensor $\overline{K}$ is objective
\begin{align}
  \label{2.16}
  \overline{K} = \overline{R}^T\Curl\overline{R} \rightarrow 
  (Q\overline{R})^T \Curl (Q\overline{R}) = 
  \overline{R}^T Q^T Q\Curl \overline{R} = \overline{K}\;,
\end{align}
where we used that $Q$ is a constant orthogonal matrix. Therefore, $\overline{K}$ is objective but also chiral due to (\ref{2.10}). These observations allow us to construct two additional terms which are also objective and chiral, namely
\begin{align}\label{2.17}
  L=R^T\Curl\overline{R}\qquad\text{and}\qquad M=\overline{R}^T\Curl R\;.
\end{align}
Here, again, $R$ is the orthogonal part of the polar decomposition, while $\overline{R}$ is the microrotation. In principle one could also consider the term $K=R^T\Curl R$. Now, we have a set of chiral and objective terms
\begin{align}\label{2.18}
  \{\overline{K},K,L,M,\overline{K}^T,K^T,L^T,M^T,\ldots\}\;,
\end{align}
from which we can construct a chiral energy functional. We note that a product of an odd number of chiral terms is required to preserve chirality. 

In addition, an energy functional is hemitropic if it is right-invariant under global rotations excluding inversions, i.e.~right-invariant under the elements of $\mathrm{SO}(3)$ \cite{IM2018}. Recall that right-invariance under $\mathrm{O}(3)$ is isotropy. Considering the (right) rotation $Q_2 \in \mathrm{SO}(3)$ we find that $F$ and $\overline{R}$ transforms as
\begin{equation}
  \label{2.19}
  F = R\,U \rightarrow R \, Q_2 U \;,\qquad
  \overline{R} \rightarrow \overline{R} \, Q_2 \;.
\end{equation}
This implies the following property
\begin{align}
  \label{2.21}
  \overline{K} = \overline{R}^T\Curl\overline{R} \rightarrow 
  (\overline{R}Q_2)^T \Curl (\overline{R}Q_2) = 
  Q_2^T\overline{R}^T \Curl \overline{R} Q_2= Q_2^T \overline{K} Q_2 \;.
\end{align}
Likewise, we find 
\begin{align}
  \label{2.22}
  L&=R^T\Curl\overline{R}  \rightarrow (R Q_2)^T\Curl\overline{R}Q_2 = Q_2^T L Q_2 \;,
  \\
  M&=\overline{R}^T\Curl R \rightarrow (\overline{R} Q_2)^T \Curl R Q_2 = Q_2^T M Q_2 \;,
\end{align}
and recall the standard identity $\tr(Q_2^T A Q_2) = \tr A$ for any matrix $A$.

Finally, we consider the energy functional
\begin{align}
  \label{2.23}
  V_\chi=\chi\tr\left(\overline{K}\;\overline{K}\;\overline{K} \right)\;,
  \quad \chi \in \mathbb{R} \;.
\end{align}
We note its three main properties:
\begin{itemize}
  \item $\overline{K}$ is $\mathrm{SO}(3)$-left-invariant, hence $V_\chi$ is $\mathrm{SO}(3)$-left-invariant and thus objective.
  \item $\overline{K}$ is chiral, $V_\chi$ is odd in $\overline{K}$ and hence $V_\chi$ is chiral, $V_\chi^\# = -V_\chi$.
  \item $V_\chi$ is $\mathrm{SO}(3)$-right-invariant (hemitropic) because
    \begin{align*}
      V_\chi=\chi\tr\left(\overline{K}\;\overline{K}\;\overline{K} \right) \rightarrow 
      \chi\tr\left(Q_2^T \overline{K} Q_2 Q_2^T \overline{K} Q_2 Q_2^T \overline{K} Q_2\right) =
      \chi\tr\left(Q_2^T \overline{K}\;\overline{K}\;\overline{K} Q_2\right) = V_\chi
    \end{align*}
    due to the cyclic property of the trace.
\end{itemize}
This means $V_\chi$ is objective, hemitropic and chiral.

Clearly, following the same approach one can consider similarly structured terms based on cubic combinations of $\overline{K}$, $L$ and $M$. For simplicity, we will focus on the easiest of these terms. The use of the dislocation density tensor (\ref{2.16}) should not be construed to imply that chirality needs a dislocated solid. Other, third order curvature measures not related to dislocations could be used instead.

\section{Chiral equations of motion in the plane}
\label{sec:eom}

The chiral energy term $V_\chi$ is by no means guaranteed to lead to a non-vanishing contribution to the equations of motion when the planar problem is considered. However, it will turn out to produce the necessary terms to preserve planar chirality.

\subsection{Variations of energy functional}

In the following we will define an energy functional from which we will derive equations of motion. After constructing an explicit solution we will discuss the chiral properties of the solution. Let us begin by defining the total energy functional including a chiral term $V_{\chi}$ of the form (\ref{2.23}) for the Cosserat material which is given by
\begin{align}
  \label{3.1}
  V = V_{\text{elastic}}(F,\overline{R}) + V_{\text{curvature}}(\overline{R}) +
  V_\chi(\overline{R}) - V_{\text{kinetic}}(\mathbf{u},\overline{R})\;.
\end{align}
where the individual terms are 
\begin{align}
  \label{3.2a}
  V_{\text{elastic}}(F,\overline{R}) &=
  \mu \left\| \sym (\overline{R}^{T}F-\id) \right\|^{2} +
  \frac{\lambda}{2}\tr\left(\sym\left(\overline{R}^{T}F-\id\right)\right)^2\;,\\
  \label{3.2b}
  V_{\text{curvature}}(\overline{R})&=
  \kappa_{1}\left\|\dev \sym (\overline{R}^{T}\;\Curl\overline{R})\right\|^{2}+
  \kappa_{2}\left\|\skewp (\overline{R}^{T}\Curl\overline{R})\right\|^{2} +
  \kappa_{3}\tr(\overline{R}^{T}\Curl\overline{R})^{2}\;,\\
  \label{3.2c}
  V_{\chi}(\overline{R}) &= \chi \tr\left(\overline{K}\;\overline{K}\;\overline{K} \right)\;,\\
  \label{3.2d}
  V_{\text{kinetic}} &= \frac{1}{2}\rho\|\dot{\mathbf{u}}\|^2 + \rho_{\text{rot}} \|\dot{\overline{R}}\|^2\;.
\end{align}
In the above we used the notation $\skewp X = (X-X^T)/2$ and $\sym X = (X+X^T)/2$ for the skew-symmetric and symmetric parts of the matrix $X$, respectively. Moreover $\dev X = X - \frac{1}{3}\tr(X) \id$ stands for the deviatoric or trace-free part of $X$.

The variation of the total energy functional
\begin{align}\label{3.5}
  \delta V = \delta V_{\text{elastic}}(F,\overline{R})+\delta V_{\text{curvature}}(\overline{R})+
  \delta V_\chi(\overline{R})-\delta V_{\text{kinetic}}(\mathbf{u},\overline{R})
\end{align}
will lead to the equations of motion by collecting corresponding terms of $\delta\overline{R}$ and $\delta F$. The detailed calculations for varying functionals $V_{\text{elastic}}$ and $V_{\text{curvature}}$ can be found in~\cite{CB2019-1}. We recall
\begin{multline}
  \delta V_{\text{elastic}}(F,\overline{R})=\Big[\mu(\overline{R}F^{T}\overline{R}+F)-
    \Big(2\mu+3\lambda\Big)\overline{R}+ \lambda \tr(\overline{R}^{T}F)\overline{R}\Big]:\delta F\\[1ex] +
  \Big[\mu F\overline{R}^{T}F-\Big(2\mu+3\lambda\Big)F+\lambda\tr(\overline{R}^{T}F)F\Big]:
  \delta \overline{R} \;.
  \label{3.6}
\end{multline}
The variations of the curvature term are given by
\begin{align}
  \delta V_{\text{curvature}}(\overline{R}) = {} &
  \Big[(\kappa_{1}-\kappa_{2})\Big((\Curl\overline{R})\overline{R}^{T}(\Curl(\overline{R}))+
    \Curl\Big[\overline{R}(\Curl\overline{R})^{T}\overline{R}\Big]\Big)
    \nonumber \\[1ex] &+
    (\kappa_{1}+\kappa_{2})\Curl\Big[\Curl\overline{R}\Big] -
    \Big(\frac{\kappa_{1}}{3}-\kappa_{3}\Big)
    \Big(4\tr(\overline{R}^{T}\Curl\overline{R})\Curl(\overline{R})
    \nonumber \\[1ex] &-
    2\overline{R}\Big[\text{grad}\Big(\tr[\overline{R}^{T}\Curl\overline{R}]\Big)\Big]^{\star}\Big)
    \Big]:\delta\overline{R}\;,
    \label{3.7}
\end{align}
where $\big(\grad \tr(A)\big)^{\star}_{ik} = \epsilon_{ijk} \partial_j \tr(A)$. For $V_\chi(\overline{R})$ we have (see Appendix~\ref{app:vef})
\begin{align}
  \delta\chi\tr\left(\overline{K}\;\overline{K}\;\overline{K}\right)=3\chi\Bigl[(\Curl\overline{R})(\overline{K}^2)+\Curl[\overline{R}(\overline{K}^2)^T]\Bigr]:\delta\overline{R}\;.
\label{3.8}
\end{align}
Collecting $\delta F$-terms gives
\begin{align}
  \label{3.8-1}
  \mu(\overline{R}F^{T}\overline{R}+F)-(2\mu+3\lambda)\overline{R}+ \lambda \tr(\overline{R}^{T}F)\overline{R}
  =: \B{A} =
  \begin{pmatrix}
    A_{11}&A_{12}&A_{13}\\
    A_{21}&A_{22}&A_{23}\\
    A_{31}&A_{32}&A_{33}
  \end{pmatrix} \;.
\end{align}
As all equations are matrix valued, we are trying to collect the relevant terms efficiently.

Starting with $F = \id + \nabla \mathbf{u}$ we find that variations of the deformation gradient tensor $\delta F$ are related to gradients of displacement $\delta F = \delta(\nabla \mathbf{u})$, consequently one needs to integrate by parts to find the variations with respect to $\delta \mathbf{u}$. More explicitly, for any matrix $\B{A}$, we have
\begin{align}
  \B{A}:\delta F=A_{ij}\delta F_{ij}\longrightarrow-\partial_jA_{ij}\delta u_i=-\left(\partial_1A_{31}+\partial_2A_{32}+\partial_3A_{33}\right)\delta\psi\;,
\end{align}
where a total derivative term was neglected.

Likewise, collecting $\delta\overline{R}$-terms from these variations can be summarised to a single matrix $\B{B}$ as shown in the following
\begin{multline}
  \label{3.9}
  3\chi\Bigl[(\Curl\overline{R})(\overline{K}^2)+\Curl[\overline{R}(\overline{K}^2)^T]\Bigr]
  +\mu F\overline{R}^{T}F-\Big(2\mu+3\lambda\Big)F+\lambda\tr(\overline{R}^{T}F)F\\
  +(\kappa_{1}-\kappa_{2})\Big((\Curl\overline{R})\overline{R}^{T}(\Curl(\overline{R}))+\Curl\Big[\overline{R}(\Curl\overline{R})^{T}\overline{R}\Big]\Big) \\[1ex]
  +(\kappa_{1}+\kappa_{2})\Curl\Big[\Curl\overline{R}\Big]-\Big(\frac{\kappa_{1}}{3}-\kappa_{3}\Big)\Big(4\tr(\overline{R}^{T}\Curl\overline{R})\Curl(\overline{R})\\[1ex]
  -2\overline{R}\Big[\text{grad}\Big(\tr[\overline{R}^{T}\Curl\overline{R}]\Big)\Big]^{\star}\Big)+2\rho_{\text{rot}} \ddot{\overline{R}}
  =: \B{B} \,.
\end{multline}

\subsection{Equations of motion and solutions}

In order to study the equations of motion in the plane, we will make the following simplifying assumptions which proved particularly useful in~\cite{CB2016-2,CB2019-1}. We assume:
\begin{enumerate}
\item Material points can only experience rotations about one axis, the $z$-axis $(0,0,1)^T$, say.
\item Displacements occur along this axis of rotation.
\item Elastic and rotational waves are both longitudinal and propagate with same wave speed $v$.
\end{enumerate}

Consequently, the microrotation matrix can be written as 
\begin{align}
  \label{3.11}
  \overline{R}=
  \begin{pmatrix}
    \cos\phi(z,t)&-\sin\phi(z,t) &0\\
    \sin\phi(z,t)&\cos\phi(z,t) &0\\
    0&0&1
  \end{pmatrix} \;.
\end{align}
The displacement vector and deformation gradient tensor take the respective forms
\begin{align}
  \label{3.12}
  u=
  \begin{pmatrix}
    0\\
    0\\
    \psi(z,t)
  \end{pmatrix}, \qquad
  F=
  \begin{pmatrix}
    1&0&0\\
    0&1&0\\
    0&0&1+\partial_{z}\psi(z,t)
  \end{pmatrix}
\end{align}
which imply $F=U$ and $\polar(F)=R=\id$. From these, we can write the explicit forms of $\overline{K}$, 
\begin{align}\label{3.12-1}
  \overline{K}=
  \begin{pmatrix}
    \partial_z\phi&0&0\\
    0&\partial_z\phi&0\\
    0&0&0
  \end{pmatrix}
  \;.
\end{align}
Now, under these assumptions only the $A_{33}$-term contributes to Eq.~(\ref{3.8-1}) which is given by
\begin{align}
  A_{33}=2\lambda\,(\cos\phi-1)+(\lambda+2\mu)\,\partial_z\psi\;.
\end{align}
This gives the equation of motion of $\psi$, it reads
\begin{align}
  \rho\,\partial_{tt}\psi+2\lambda\sin\phi\;\partial_z\phi-(\lambda+2\mu)\partial_{zz}\psi=0\;.
  \label{eom:psi}
\end{align}

On the other hand the quantity $\mathbf{B}:\delta\overline{R}$ becomes
\begin{align}\label{3.13}
  \mathbf{B}:\delta\overline{R}&=
  \tr[B^T\delta\overline{R}]=
  \tr\left[\begin{pmatrix}B_{11}&-B_{12}&0\\B_{12}&B_{11}&0\\0&0&B_{33}\end{pmatrix}
    \begin{pmatrix}-\sin\phi&-\cos\phi&0\\\cos\phi&-\sin\phi&0\\0&0&0\end{pmatrix}\right]\delta\phi
  \nonumber \\
  &=-(2B_{11}\sin\phi+2B_{12}\cos\phi)\delta\phi\;.
\end{align}
The required components of $\B{B}$ are given by
\begin{align}
  \label{3.10}
  B_{11}= {}&-2(\lambda+\mu)+\frac{1}{3}\cos\phi\Bigl[3(2\lambda+\mu)-6\rho_{\text{rot}}(\partial_t\phi)^2+(\partial_z\phi)^2(\kappa_1-3\kappa_2+24\kappa_3+18\chi\partial_z\phi)\Bigr]
  \nonumber \\
  &+\lambda\partial_z\psi+\frac{2}{3}\sin\phi\Bigl[-3\rho_{\text{rot}}\partial_{tt}\phi+(\kappa_1+6\kappa_3+9\chi\partial_z\phi)\partial_{zz}\phi\Bigr] \;,\\
  B_{12}= {}&\frac{1}{3}\sin\phi\Bigl[(3\mu+6\rho_{\text{rot}}(\partial_t\phi)^2-(\partial_z\phi)^2(\kappa_1-3\kappa_2+24\kappa_3+18\chi\partial_z\phi)\Bigr]
  \nonumber \\
  &+\frac{2}{3}\cos\phi\Bigl[-3\rho_{\text{rot}}\partial_{tt}\phi+(\kappa_1+6\kappa_3+9\chi\partial_z\phi)\partial_{zz}\phi\Bigr] \;.
\end{align}
Putting everything together gives the equation of motion for $\phi$ together with the previous equation for $\psi$ (\ref{eom:psi}). This coupled system of equations is given by
\begin{align}
  \label{3.14}
  \rho_{\text{rot}}\,\partial_{tt}\phi-\left(\frac{\kappa_1+6\kappa_3}{3}\right)\partial_{zz}\phi-3\chi\partial_z\phi\;\partial_{zz}\phi+(\lambda+\mu)(1-\cos\phi)\sin\phi-\frac{\lambda}{2}\sin\phi\;\partial_z\psi=0\;, \\
  \rho\,\partial_{tt}\psi+2\lambda\sin\phi\;\partial_z\phi-(\lambda+2\mu)\partial_{zz}\psi=0\;.
  \label{eom:psiA}
\end{align}

Now, we seek solutions of the form $\phi=f(z-vt)$ and $\psi=g(z-vt)$ where $v$ is the same wave speed for both the elastic and the rotational wave propagation~\cite{CB2016-2,CB2019-1}. This means they both satisfy the wave equation $\partial_{tt}f=v^2\partial_{zz}f$. We introduce the notation $s=z-vt$ and denote differentiation with respect to $s$ by a prime. 

Putting this ansatz into the equation of motion of $\psi$ (\ref{eom:psiA}) gives
\begin{align}
  \rho v^2 g'' + 2\lambda \sin(f) f' - (\lambda + 2\mu) g'' &= 0
  \nonumber \\ \quad \Leftrightarrow \quad
  (\rho v^2 - (\lambda + 2\mu)) g'' = -2\lambda \sin(f) f' &= 2\lambda \frac{d}{ds}(\cos(f)) \;.
\end{align}
Consequently, we can now integrate with respect to $s$ which gives
\begin{align}
  g' = \frac{2\lambda}{\rho v^2-(\lambda+2\mu)}\cos(f)+C_1\;,
\end{align}
for some constant of integration $C_1$. Since $\partial_z \psi = g'$ we can now eliminate this term from the equation of motion (\ref{3.14}) which becomes
\begin{multline}
  \rho_{\text{rot}} v^2 f'' -\left(\frac{\kappa_1+6\kappa_3}{3}\right) f'' - 3\chi f' f'' +
  (\lambda+\mu)(1-\cos\phi)\sin\phi \\
  -\frac{\lambda}{2}\sin(f)\left(\frac{2\lambda}{\rho v^2-(\lambda+2\mu)}\cos(f)+C_1\right) =0\;.
\end{multline}
After rearranging terms this can be rewritten as follows
\begin{multline}
  \rho_{\text{rot}} v^2 f'' -\left(\frac{\kappa_1+6\kappa_3}{3}\right) f'' - 3\chi f' f'' +
  \left[(\lambda+\mu) -\frac{\lambda}{2} C_1 \right]\sin(f) \\
  -\frac{1}{2}\left[(\lambda+\mu) + \frac{\lambda^2}{\rho v^2-(\lambda+2\mu)}\right]\sin(2f) =0\;.
  \label{3.15}
\end{multline}
When setting $\chi = 0$ this becomes the \textit{double sine-Gordon equation} \cite{AD2003} which we studied previously in~\cite{CB2019-1} where a soliton-like solution could be constructed. When $\chi \neq 0$ there is an additional non-linearity of the form $\chi f' f''$.

\subsection{Constructing approximate solutions}

Inspection of equation~(\ref{3.15}) shows that each of these five terms can be integrated after the entire equation is multiplied by $f'$. This yields
\begin{multline}
  \frac{1}{2}\left[\rho_{\text{rot}} v^2 -\left(\frac{\kappa_1+6\kappa_3}{3}\right)\right] (f')^2 -
  \chi (f')^3 - \left[(\lambda+\mu) -\frac{\lambda}{2} C_1 \right]\cos(f) \\ +
  \frac{1}{4}\left[(\lambda+\mu) + \frac{\lambda^2}{\rho v^2-(\lambda+2\mu)}\right]\cos(2f) = C_2\;,
  \label{3.16}
\end{multline}
where $C_2$ is another constant of integration. Formally this is a cubic equation in $f'$ (quadratic in $f'$ if $\chi=0$) which can, in principle, be solved for $f'$ and will give three different solutions in general. The resulting equation is always of the general form $f'=H(f)$ and is hence separable. This means we have reduced finding a solution to our system of nonlinear wave equations (\ref{eom:psi}) and (\ref{3.14}) to an integration problem. Note that this solution depends on the eight parameters $\{v,\lambda,\mu,\rho,\kappa_1,\kappa_3,\rho_{\rm rot},\chi\}$ and two constants of integration $C_1$ and $C_2$. In general these solutions will involve special functions if an explicit solution can be found.

In the following we will construct an approximated solution to (\ref{3.16}) using a suitable choice of the constants of integration. The chiral parameter $\chi$ is assumed to be small so that a series expansion in this parameter can be made. We choose $C_1=2(\lambda+\mu)/\lambda$, put $\mathcal{F}(s)=2f(s)$ and choose $C_2$ such that one can write
\begin{equation}
  \label{3.17}
  (\mathcal{F}')^2-\tilde{\chi}(\mathcal{F}')^3=2m^2(1-\cos(\mathcal{F}))
\end{equation}
where the following constants were introduced
\begin{equation}
  \label{3.18}
  \tilde{\chi}=\frac{3\chi}{\rho_{\text{rot}}v^2-(\kappa_1+6\kappa_3)}\quad\text{and}\quad
  m^2=\frac{3v^2\rho(\lambda+\mu)-3\mu(3\lambda+2\mu)}{(\lambda+2\mu-v^2\rho)(\kappa_1+6\kappa_3-3v^2\rho_{\text{rot}})} \;.
\end{equation}
Now, to solve this, we use regular perturbation methods. First we write the function $\mathcal{F}(s)$ as a series expansion 
\begin{equation}
  \label{3.19}
  \mathcal{F}(s)=\mathcal{F}_0(s)+\tilde{\chi}\, \mathcal{F}_1(s)+\tilde{\chi}^2\, \mathcal{F}_2(s)+\cdots
\end{equation}
in the chiral parameter $\tilde{\chi}$ which we assume to be small $\tilde{\chi} \ll 1$. We impose the initial conditions $\mathcal{F}(0)=\mathcal{F}_0(0)=\pi$, and $\mathcal{F}_1(0)=\mathcal{F}_2(0)=\cdots=0$.

Let us begin with the function $\mathcal{F}_0(s)$ which satisfies
\begin{equation}
  \label{3.19-1}
  \frac{d\mathcal{F}_0}{ds} = \pm m \sqrt{2(1-\cos (\mathcal{F}_0))}\;,
\end{equation}
the solution of which is well known
\begin{equation}
  \label{3.20}
  \mathcal{F}_0(s)=4\arctan(e^{\pm ms})\;.
\end{equation}
The positive sign solution is known as the kink solution while the negative sign solution is the anti-kink. For nonzero $\tilde{\chi}$, we substitute $\mathcal{F}(s)$ into the equation of motion (\ref{3.17}) and make a series expansion in $\tilde{\chi}$ up to the first order (for the second order solution, see Appendix~\ref{app;order2}).

In first order in $\tilde{\chi}$ we have
\begin{equation}
  \label{3.20-2}
  -(\mathcal{F}'_0)^3-2m^2\sin(\mathcal{F}_0)\,\mathcal{F}_1+2\,\mathcal{F}'_0\,\mathcal{F}'_1 = 0\;,
\end{equation}
where $\mathcal{F}'_0=2\,m\sech(ms)=2m/\cosh(ms)=4m/(e^{ms}+e^{-ms})$ follows from (\ref{3.20}). With the initial condition $\mathcal{F}_1(0)=0$ we obtain
\begin{equation}
  \label{3.20-3}
  \mathcal{F}_1(s) = m\sech(ms)(4\arctan(e^{ms})-\pi) = \frac{1}{2}\mathcal{F}'_0(\mathcal{F}_0-\pi)\;,
\end{equation}
which gives the first order solution
\begin{equation}
  \label{3.20-4}
  \mathcal{F}(s) = 4\arctan(e^{ms})+\tilde{\chi}\,m\sech(ms)(4\arctan(e^{ms})-\pi)\;.
\end{equation}

Considering the asymptotic behaviour of the solution, it would be ideal to have the boundary conditions $\phi(-\infty,t)=\phi(+\infty,t)=0$ for all times $t$. This is because the material should be in its original state as we approach spatial infinity $z \to \pm\infty$. To incorporate these boundary conditions, we note that the kink and anti-kink solutions (\ref{3.20}) are readily applicable so that we can redefine $\mathcal{F}_0(s)$ as follows
\begin{equation}
  \label{3.20-5}
  \mathcal{F}_0(s)=
  \begin{cases}
    4\arctan(e^{+ms})\quad&\text{if $s<0$}\\
    4\arctan(e^{-ms})\quad&\text{if $s>0$}\;.
  \end{cases}
\end{equation}
Next, we can apply the same reasoning to the first order contribution (\ref{3.20-3}). Notice that the solution $\mathcal{F}_1(s)$ is invariant under the change $m\to-m$, which is the equivalent to $s\to -s$ in this case. This is clear from (\ref{3.20-3}) using the identity $\arctan(X^{-1})=\pi/2-\arctan(X)$.

Now, putting back $s=z-vt$ and $\phi(z,t)=f(z-vt)=\mathcal{F}/2$ and using (\ref{3.20-4}), the first order solution for $\phi(z,t)$ becomes
\begin{equation}
  \label{3.24}
  \phi(z,t)=
  \begin{cases}
    2\arctan(e^{+m(z-vt)}) + \frac{\tilde{\chi}\,m}{2}\sech(m(z-vt))(4\arctan(e^{m(z-vt)})-\pi)
    &\text{if $z<v\,t$}\\
    2\arctan(e^{-m(z-vt)}) + \frac{\tilde{\chi}\,m}{2}\sech(m(z-vt))(4\arctan(e^{m(z-vt)})-\pi)
    &\text{if $z>v\,t$}\;.
  \end{cases}
\end{equation}

\begin{figure}[!htb]
  \[\includegraphics[scale=0.48]{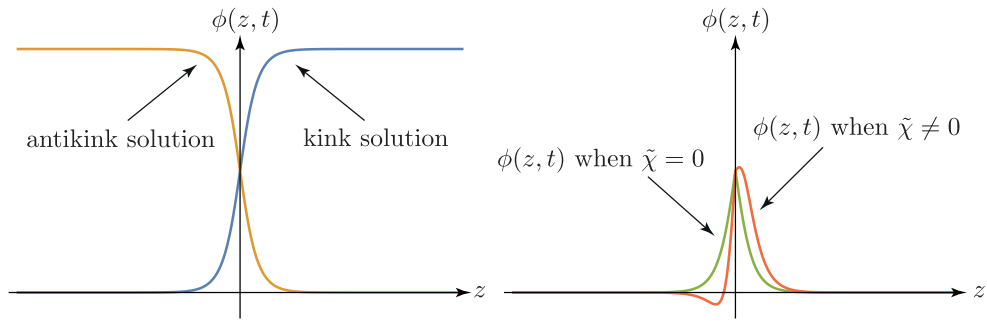}\]
  \caption{Left panel: The kink and anti-kink solutions obtained in (\ref{3.20-4}) are shown at $t=0$, indicating the asymptotic behaviours of $\phi_{\text{kink}}(+\infty,t)$ and $\phi_{\text{anti-kink}}(-\infty,t)$. Right panel: The piece-wise definition of $\phi(z,t)$ is shown for $\tilde{\chi}=0$ and $\tilde{\chi}\neq 0$.}
\label{f002}
\end{figure}

The solution constructed in this way gives rise to a localised wave which propagates along the $z$-axis. The amplitude decreases exponentially such that the material recovers its original shape once the wave passes through, see Fig.~\ref{f002}.

If we apply the inversion operator to (\ref{3.15}) it will yield a new equation of motion where $\phi(z,t)$ is replaced by $\phi^\#(z,t)=\phi(-z,t)$. Then the $\chi$-term in (\ref{3.15}) acquires an additional minus sign due to the presence of three derivatives with respect to $z$, one second derivative and one first derivative. This results in
\begin{multline}
  \left[\rho_{\text{rot}} v^2 -\left(\frac{\kappa_1+6\kappa_3}{3}\right)\right]\partial_{zz}\phi^\# + 3\chi\partial_z\phi^\#\partial_{zz}\phi^\# +
  \left[(\lambda+\mu) -\frac{\lambda}{2} C_1 \right]\sin(\phi^\#) \\
  -\frac{1}{2}\left[(\lambda+\mu) + \frac{\lambda^2}{\rho v^2-(\lambda+2\mu)}\right]\sin(2\phi^\#) =0\;.
  \label{3.25}
\end{multline}
At first sight, it is not straightforward to see that (\ref{3.25}) is the chiral counterpart of (\ref{3.15}) other than the sign change of the $\chi$ term. First of all, the wave speeds for both wave equations are identical and, formally, we will arrive at the same solution, as it should be. Recall that $\phi^\#(z,t)=\phi(-z,t)$ which means that the solution to one equation gives the solution to the other by reflection along the $z$-axis. In other words, the wave solution of (\ref{3.15}) will be the right-moving wave while the solution of (\ref{3.25}) describes the left-moving wave. Note that $\phi(z,t)$ is an even function if $\chi=0$ while this is not the case when $\chi\neq 0$.

Furthermore, the left-moving wave will be governed by the rotation matrix $\overline{R}^\#=-\overline{R}$ with respect to the inverted axis $-z$. Hence the orientations of left-moving wave and right-moving wave for the material elements, which experience microrotations, are identical but reflected. Therefore, we can conclude that two wave solutions are chiral to each other. The right-moving waves `rotates' anti-clockwise while the left-moving one `rotates' clockwise.

\begin{figure}[!htb]
  \centering
  \includegraphics[width=0.48\textwidth]{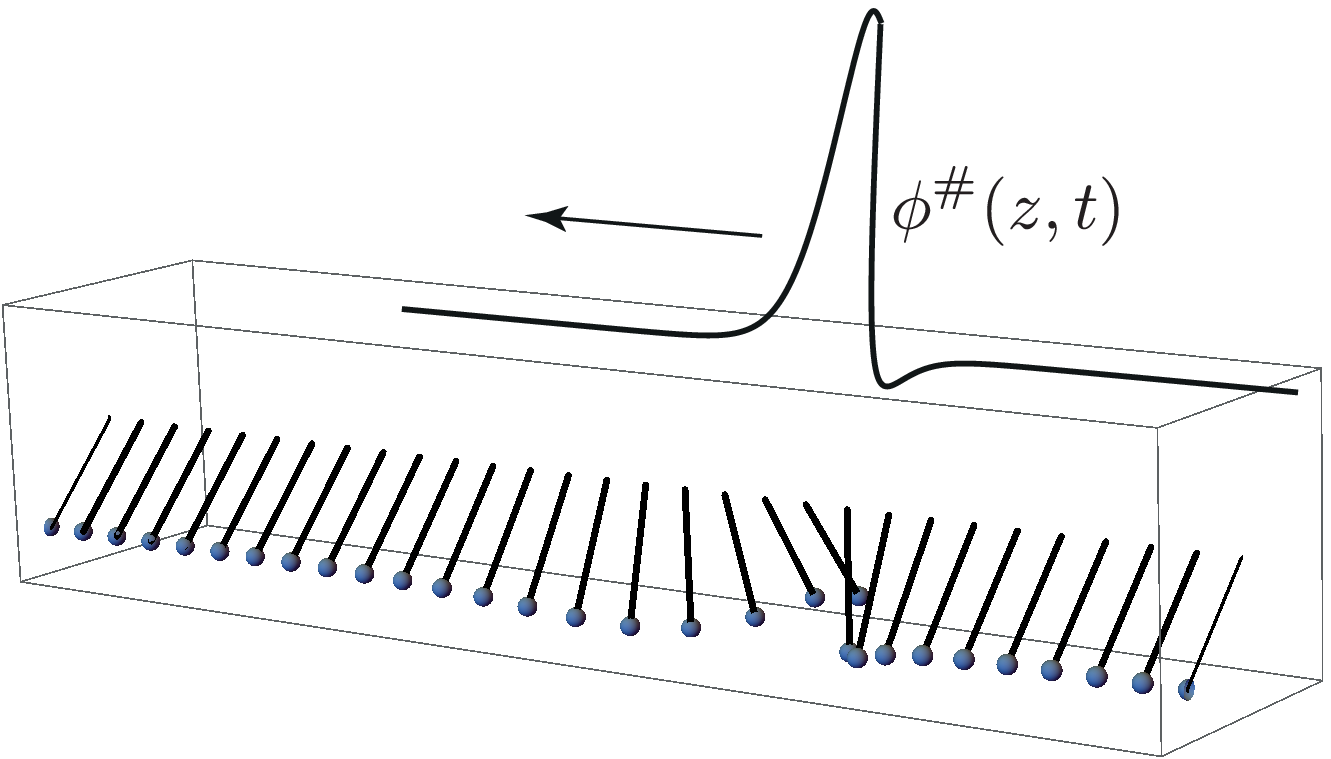}
  \hfill
  \includegraphics[width=0.48\textwidth]{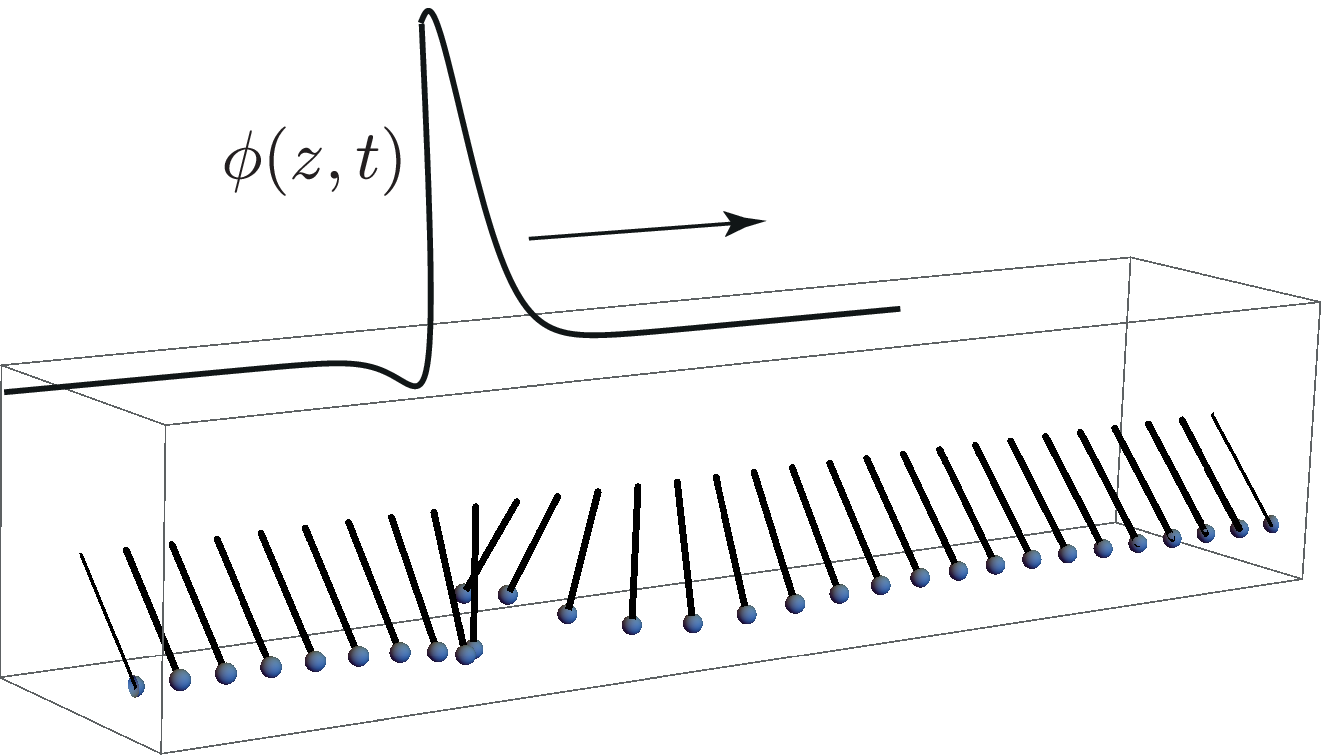}
  %\parbox{3in}{\includegraphics[scale=0.5]{./figure/004}}
  %\parbox{3in}{\includegraphics[scale=0.5]{./figure/005}}
  \caption{Right panel: The right-moving wave shows the rotational deformations governed by the microrotation $\overline{R}$ on the material elements, modelled by pendulums, as the wave propagates to the right along the $z$-axis. Left panel: The left-moving wave gives the same rotational defects with $\overline{R}^\#$ as it propagates to the left. We set $\tilde{\chi}=0.6$ and $m=2$.}
\label{f003}
\end{figure}

We emphasise that this inversion operation on $\overline{R}$ only takes effect on the terms coupled to the chiral part. Specifically, only the chiral energy $V_\chi(\overline{R})$ is responsible for generating the chiral counterpart $\overline{R}^\#$ while the other energy functional terms are invariant under this inversion operation. This is also evident from the variation with respect to $\delta\overline{R}$ in (\ref{3.9}). Consequently, the chiral coupled term in the equation of motion and its solutions experience the effects of the inversion operation on the rotational matrix as the wave propagate along the axes. The travelling wave solutions are shown in Fig.~\ref{f003}.

\section{Discussion}
\label{sec:dis}

We started by defining the inversion operator $\#$ in three-dimensional space to define what is meant by chirality: in simple words an energy term which acquires an additional minus sign when evaluated in the inverted coordinates. By recognising the fact that it is possible to construct chiral terms which are objective and hemitropic we suggested various chiral energy functionals and formulated a total energy functional. This consisted of an elastic energy and a curvature energy functional accompanied by a chiral term. After formulating the equations of motion in full generality a planar ansatz was used to study the equations of motion in this planar setting. These equations retained information about chirality and we were able to present an approximated solution assuming the chirality parameter to be small. We emphasise that we have used a chiral energy functional based on a simple construction, but one may use more complicated chiral energies by considering the terms given in the set (\ref{2.18}).

An interesting question to consider in general is whether the chiral energy $V_\chi$ should change its sign under inversions or not. If it does, as we assumed in the above, one of the two chiral states has a lower energy state than the other. Consequently one of the two chiral states is favoured energetically. This is somewhat at odds with the fact that simple chiral molecules appear in equal proportion in chemical experiments. On the other hand, there are also known examples, see again~\cite{SS2009}, where the chiral counterpart is unstable hence justifying a lower energy for one of the two states. Our proposed model is able to take into account both setups by simply setting either $\chi^\# = \chi$ or $\chi^\# = -\chi$. The case $\chi^\# = \chi$ corresponds to the chiral parts having a different energy states, thereby modelling a scenario where one of the chiral states is unstable while the other one is stable, for instance. When working with $\chi^\# = -\chi$ the energy of both chiral states is identical. Our solution is valid for both types of models as the value of the chiral parameter only affects the shape of the localised solution but does not change any of the interpretation. 

\subsection*{Acknowledgement}

Yongjo Lee is supported by EPSRC Doctoral Training Programme (EP/N509577/1).

\footnotesize
\appendix

\section{The rotation matrix under inversion}
\label{app:rot}

Consider the transformation from the reference configuration $X_L$ to the spatial configuration $x_i=R_{iL}X_L$ by a rotation
\begin{align}
  \label{a1}
  R_{iL}=\delta_{iL}\cos\phi+\epsilon_{ijL}n_j\sin\phi+(1-\cos\phi)n_in_L
\end{align}
where $n_i$ is an axis of rotation and $\phi$ is the rotation angle in spatial configuration. $n_L$ is an axial vector in the reference configuration. Then applying the inversion operator $\#$ gives
\begin{align}
  \label{a2}
  R^\#_{iL}=-\delta_{iL}\cos\phi-\epsilon_{ijL}n_j\sin\phi-(1-\cos\phi)n_in_L
\end{align}
in which $\delta_{iL}=\partial x_i/\partial X_L$ acquires the negative sign and $n_i$ changes its direction under the inverted coordinates but $n_L$ remains unaffected in the reference configuration. This agrees with result (\ref{2.5}) that $R^\#_{iL}=-R_{iL}$. More generally speaking, for $R\in \mathrm{O}(3)$, we have $\det R=\pm 1$ and we can write
\begin{align}
  \label{a3}
  R=s\,e^A=
  \begin{cases}
    \det R=+1\;&\text{if $s=+1$}\\
    \det R=-1\;&\text{if $s=-1$}
  \end{cases}
\end{align}
where $A$ is a $3\times 3$ skew-symmetric matrix. This classification identifies the relation between $R\in \mathrm{SO}(3)$ and its chiral counterpart $R^\#$. If $\det R=s=+1$ we have $\det R^\# = -1$ and vice versa.

On an equal footing, the deformation gradient tensor is defined by $F_{iR}=\frac{\partial x_i}{\partial X_R}$ for $x_i=X_i+u_i(X_k)$ and applying the operator $\#$ gives
\begin{align}
  \label{a4}
  F^\#_{iR}=\frac{\partial(-x_i)}{\partial X_R}=-F_{iR}\;,
\end{align}
also in agreement with (\ref{2.4}).

\section{Variations of the energy functional}
\label{app:vef}
We vary the chiral energy functional $V_\chi=\chi\tr(\overline{K}^3)$ as follows
\begin{align}
  \delta\chi\tr(\overline{K}^3)&=3\chi(\overline{K}^2)^T:\delta\overline{K} =
  3\chi\Bigl[(\overline{K}^2)^T_{ij}\delta(\overline{R}^T\Curl\overline{R})_{ij}\Bigr]=
  3\chi\Bigl[(\overline{K}^2)^T_{ij}\left(\delta\overline{R}_{mi}(\Curl\overline{R})_{mj}+\overline{R}_{mi}(\delta\Curl\overline{R})_{mj}\right)\Bigr]
  \nonumber \\
  &=3\chi\Bigl[(\Curl\overline{R})_{mj}(\overline{K}^2)^T_{ij}\delta\overline{R}_{mi}+\overline{R}_{mi}(\overline{K}^2)^T_{ij}(\delta\Curl\overline{R})_{mj}\Bigr]
  \nonumber \\
  &=3\chi\Bigl[(\Curl\overline{R})(\overline{K}^2):\delta\overline{R}+\overline{R}(\overline{K}^2)^T:\delta\Curl\overline{R}\Bigr]
  =3\chi\Bigl[(\Curl\overline{R})(\overline{K}^2):\delta\overline{R}+\Curl[\overline{R}(\overline{K}^2)^T]:\delta\overline{R}\Bigr]
  \nonumber \\
  &=3\chi\Bigl[(\Curl\overline{R})(\overline{K}^2)+\Curl[\overline{R}(\overline{K}^2)^T]\Bigr]:\delta\overline{R}\;.
\end{align}

\section{Second order approximation}
\label{app;order2}
By following the steps (\ref{3.19})--(\ref{3.20-3}) one finds $\mathcal{F}_2(s)$ satisfies the differential equation
\begin{equation}
  -m^2\cos(\mathcal{F}_0)\,(\mathcal{F}_1)^2-2m^2\sin(\mathcal{F}_0)\,\mathcal{F}_2-3(\mathcal{F}'_0)^2\mathcal{F}'_1+(\mathcal{F}'_1)^2+2\mathcal{F}'_0\,\mathcal{F}'_2 = 0 \;.
\end{equation}
This depends on the lower order solutions $\mathcal{F}_0$ and $\mathcal{F}_1$. We can solve this by using the already obtained results for $\mathcal{F}_0$ and $\mathcal{F}_1$, with the initial condition $\mathcal{F}_2(0)=0$. The solution is given by
\begin{equation}
\mathcal{F}_2(s)=\frac{1}{8}\mathcal{F}''_0\left[\left(\mathcal{F}_0+\frac{(\mathcal{F}'_0)^2}{\mathcal{F}''_0}-\pi\right)^2-12-\left(\frac{(\mathcal{F}'_0)^2}{\mathcal{F}''_0}\right)^2\right] \;.
\end{equation}

\end{document}